\documentclass[11pt]{article}
\usepackage{amsmath}
\usepackage{amsfonts,amssymb,latexsym}

\begin{document}

\newcommand{\nl}{\nonumber\\}
\newcommand{\nnl}{\nl[6mm]}
\newcommand{\nle}{\nl[-2.5mm]\\[-2.5mm]}
\newcommand{\nlb}[1]{\nl[-2.0mm]\label{#1}\\[-2.0mm]}

\renewcommand{\leq}{\leqslant}
\renewcommand{\geq}{\geqslant}

\newcommand{\be}{\bes}
\newcommand{\ee}{\ees}
\newcommand{\bes}{\begin{eqnarray}}
\newcommand{\ees}{\end{eqnarray}}
\newcommand{\eens}{\nonumber\end{eqnarray}}

\renewcommand{\/}{\over}
\renewcommand{\d}{\partial}
\newcommand{\e}{\mathrm e}
\newcommand{\no}[1]{{\,:\kern-0.7mm #1\kern-1.2mm:\,}}

\newcommand{\ket}[1]{\big{|}#1\big{\rangle}}

\newcommand{\vect}{{\mathfrak{vect}}}

\newcommand{\eps}{\epsilon}
\newcommand{\ww}{\omega}
\newcommand{\Tot}{{\mathrm Tot}}

\newcommand{\xx}{{\mathbf x}}
\newcommand{\mm}{{\mathbf m}}
\newcommand{\qq}{{\mathbf q}}

\newcommand{\hatj}{\hat\jmath}
\newcommand{\cm}{{,\mm}}
\newcommand{\cmj}{{,\mm+\hatj}}
\newcommand{\cmnoll}{{,\mm+\hat0}}

\newcommand{\EE}{{\mathcal E}}
\newcommand{\RR}{{\mathbb R}}

\title{{Quantum Jet Theory, Observer Dependence, and Multi-dimen\-sional 
Virasoro algebra}}

\author{T. A. Larsson \\
Vanadisv\"agen 29, S-113 23 Stockholm, Sweden\\
email: thomas.larsson@hdd.se}

\maketitle
\begin{abstract}
We review some key features of Quantum Jet Theory: observer 
dependence, multi-dimen\-sional Virasoro algebra, and the prediction 
that spacetime has four dimen\-sions.
\end{abstract}

\smallskip

\section{Introduction}

Every experiment is an interaction between a system and an observer,
and the result depends on the physical properties of both. In
particular, a real observation depends on the observer's mass $M$ and
charge $e$. QFT predictions are independent of these quantities, which
means that some tacit assumptions have been made: the observer's charge is
small (so the observer does not disturb the system) and his mass is
large (so he follows a well-defined classical trajectory in
spacetime; in particular, the observer's position and velocity at
equal times commute). This assumption is unproblematic for all
interactions except gravity, where charge and mass are the same; heavy
mass equals inert mass.

The problem with quantum gravity is thus that an obese observer will
collapse into a black hole of his own, whereas a skinny observer can
not know where he is (he can only know where he was). From this 
perspective, assuming that their observations are identical becomes 
absurd. This is the
simple physical reason why QFT must be incompatible with gravity.
Alas, it also suggests a route to remedy the problem: make the
observer into a physical entity with quantum dynamics. The predictions
of such a theory will depend on $M$ and $e$, and it must reduce to
QFT in the limit $G = 0$, $M \to \infty$, $e \to 0$, 
and to general relativity in the limit $\hbar = 0$, $M \to 0$.

To describe observer-dependent physics we need observer-dependent
mathematics. Fortunately, such mathematics is available in the form of
Taylor expansions, or jets\footnote{Locally, a $p$-jet can be uniquely
coordinatized by a Taylor expansion truncated at order $p$.}. Namely,
a Taylor series does not only depend on the function being expanded,
but also on the choice of expansion point, i.e. the observer's
position. This motivates the name Quantum Jet Theory (QJT).

Despite the simplicity of the argument above, this is not historically
how QJT was discovered. Instead, the story started over twenty years
ago when I made two observations:
\begin{itemize}
\item
The symmetry of general relativity is the algebra of spacetime 
diffeomorphisms.
\item
CFT tells us that all interesting (local, unitary) quantum
represen\-tations of the diffeomorphism algebra on the circle are
anomalous, i.e. represen\-tations of the Virasoro algebra.
\end{itemize}
Putting these facts next to each other, it became obvious that
the correct symmetry of quantum gravity must be some multi-dimen\-sional
generalization of the Virasoro algebra. This algebra was subsequently
discovered \cite{Lar91,RM94}. and its off-shell represen\-tations were
understood in \cite{Lar98}. Unlike the classical represen\-tations,
which act on tensor fields, the quantum represen\-tations do not act on
fields, but on spacetime histories of tensor-valued $p$-jets. This
is the link between the multi-dimen\-sional Virasoro algebra,
QJT, and observer dependence.

An extension of the diffeomorphism algebra is a gauge anomaly, which
according to standard wisdom is inconsistent. However, this is not
necessarily true. A gauge anomaly turns a classical gauge symmetry
into a quantum global symmetry, which acts on the Hilbert space rather
than reducing it. The quantum theory may or may not be inconsistent,
depending on whether the action of the anomalous gauge symmetry is
unitary or not. A toy example of a consistent theory with an anomalous
gauge symmetry is the subcritical free string, which according to the
no-ghost theorem can be quantized with a ghost-free spectrum, despite
its conformal anomaly \cite{GSW87}.

In this review we follow the historical path to QJT. We start with the
multi-dimen\-sional Virasoro algebra in section \ref{sec:mVir} and
construct its lowest-energy represen\-tations in section \ref{sec:ler}.
In section \ref{sec:reg} we treat QJT as a regularization, and show
how the $p$-jet phase space becomes infinite-dimen\-sional because the
equations of motion are undefined on the ``skin''. Finally we indicate
in section \ref{sec:fin} how the divergent parts of anomalies can be
cancelled. For a realistic choice of field content this uniquely
singles out four spacetime dimen\-sions.

\section{ Multi-dimen\-sional Virasoro algebra }
\label{sec:mVir}

The Virasoro algebra,
\be
[L_m, L_n] = (n-m)L_{m+n} - {c\/12} (m^3-m) \delta_{m+n},
\label{Vir}
\ee
where $\delta_m$ is the Kronecker delta, is the unique central
extension of the algebra $\vect(1)$ of vector fields (or infinitesimal
diffeomorphisms) in one dimen\-sion. We want to find analogous
extensions of $\vect(d)$, the algebra of vector fields in $d$
dimen\-sions. Taken at face value, the prospects to succeed appear 
bleak, due to two no-go theorems:
\begin{itemize}
\item
$\vect(d)$ does not possess any central extension when $d > 1$.
\item
In QFT, there are no diff anomalies in four dimen\-sions \cite{Bon86}.
\end{itemize}
However, no theorem is stronger than its axioms. If we relax some
assumptions above, the no-go theorems can be evaded and a
multi-dimen\-sional Virasoro algebra can be constructed. The crucial
assumptions are encoded in the keywords ``central'' and ``in QFT''.
\begin{itemize}
\item
$\vect(d)$ does not possess any {\em central} extension when $d > 1$,
but it does possess {\em non-central} extensions which nevertheless
reduce to the usual, central, Virasoro extension when $d = 1$. In
general, we construct extensions by the module of closed
$(d-1)$-forms. When $d=1$, a closed zero-form is a constant function,
and the extension is central. When $d > 1$, a closed $(d-1)$-form
transforms in a nontrivial way under diffeomorphisms, but
nontrivial Lie algebra extensions still exist.
\item
The construction of diff anomalies in four dimen\-sions is not 
possible within the framework of OFT, but it can be done in QJT.
\end{itemize}

To make the connection to the ordinary Virasoro algebra very explicit,
it is instructive to write down the brackets in a Fourier basis. It is
clear that the Virasoro algebra (\ref{Vir}) can be written in the form
\bes
[L_m, L_n] &=& (n-m)L_{m+n} + c m^2 n S_{m+n}, \nl
{[}L_m, S_n] &=& (n+m)S_{m+n}, 
\nlb{nVir}
{[}S_m, S_n] &=& 0, \nl
m S_m &=& 0.
\eens

It is easy to see that this form of the Virasoro algebra is equivalent
to (\ref{Vir}), apart from a linear cocycle that has been absorbed
into a redefinition of $L_0$. The new formulation (\ref{nVir})
immediately generalizes to $d$ dimen\-sions. The generators
$L_\mu(m) = -i \exp(i m_\nu x^\nu) \d_\mu$
and $S^\mu(m)$ satisfy the relations
\bes
[L_\mu(m), L_\nu(n)] &=& n_\mu L_\nu(m+n) - m_\nu L_\mu(m+n) \nl 
&&+\ (c_1 m_\nu n_\mu + c_2 m_\mu n_\nu) m_\rho S^\rho(m+n), \nl
{[}L_\mu(m), S^\nu(n)] &=& n_\mu S^\nu(m+n)
 + \delta^\nu_\mu m_\rho S^\rho(m+n), 
\nlb{mVir}
{[}S^\mu(m), S^\nu(n)] &=& 0, \nl
m_\mu S^\mu(m) &=& 0.
\eens
This is an extension of $\vect(d)$ by the abelian ideal with basis
$S^\mu(m)$. Geometrically, we can think of $L_\mu(m)$ as a vector
field and
$S^\mu(m) = \allowbreak \eps^{\mu\nu_2..\nu_d}
\allowbreak S_{\nu_2..\nu_d}(m)$ 
as a dual one-form (and $S_{\nu_2..\nu_d}(m)$ as an $(d-1)$-form); the
last condition expresses closedness.

The cocycle proportional to $c_1$ was discovered by 
Rao and Moody \cite{RM94}, and the one proportional to $c_2$ by
myself \cite{Lar91}. There is also a similar multi-dimen\-sional 
generalization of affine Kac-Moody algebras.
The multi-dimen\-sional Virasoro and 
affine algebras are often refered to as ``Toroidal Lie algebras''
in the mathematics literature.

\section{ Lowest-energy represen\-tations and QJT }
\label{sec:ler}

In the previous section we constructed Virasoro-like extensions of 
$\vect(d)$, thus evading the first no-go theorem. This
leaves two major problems: how to build represen\-tations, and how to
avoid the fact that there are no diff anomalies in 4D in QFT. The
solution to both problems is the same: the represen\-tations act on
trajectories in jet space, to which the no-go theorem does not apply.

Let us compare with the ordinary Virasoro algebra, whose
represen\-tation theory may be viewed as QFT. Building Fock
represen\-tations consists of three steps:
\begin{enumerate}
\item
Start from a classical represen\-tation, which acts on primary fields,
i.e. scalar densities.
\item
Add canonical momenta.
\item
Normal order.
\end{enumerate}
Na\"\i vely, one could expect to build a QFT represen\-tation of 
$\vect(d)$ in the same manner, except that we could choose any
tensor density as the starting point. However, this strategy fails, for
several reasons:
\begin{enumerate}
\item
To define the Fock vacuum we must single out a privileged time or
energy direction, which is unconfortable when studying spacetime
diffeomorphisms.
\item
Normal ordering of bilinears always leads to a central extension, but
the Virasoro extension is noncentral when $d > 1$.
\item
Irremovable infinitities arise even after normal ordering, making the
approach useless.
\end{enumerate}
These problems have proved unsurmountable, and lead to the conclusion
that it is impossible to construct nontrivial lowest-energy 
represen\-tations of $\vect(d)$ within the framework of QFT.

The resolution of this paradox appeared in the seminal work by Rao and
Moody \cite{RM94}. In the physics-flavored language of \cite{Lar98},
their construction can be described as follows:
\begin{enumerate}
\item
Start from a classical realization acting on trajectories
in $p$-jet space, instead of a represen\-tation acting on fields.
\item
Add canonical momenta for the jets.
\item
Normal order.
\end{enumerate}
The problem with infinities is avoided because a $p$-jet only has
finitely many degrees of freedom, and a trajectory in $p$-jet space
thus consists of finitely many functions of a single variable. This
is precisely the situation where normal ordering works without 
producing infinities.

Locally, a $p$-jet is essentially the same thing as a Taylor
expansion; a $p$-jet has a unique representative which is a polynomial
of order at most $p$, which is the Taylor series truncated at order
$p$. A $p$-truncated Taylor series around the point $q = (q^\mu)$ 
takes the form
\be
\phi(x) = \sum_{|m|\leq p} {1\/m!} \phi_{,m} (x - q)^m.
\label{pjet}
\ee
This formula is written in a form which may appear one-dimen\-sional,
but with standard multi-index notation it makes sense also for
$d > 1$. E.g., $m = (m_0, m_1, ..., m_{d-1})$ is a multi-index with 
length $|m| = \sum_{\mu=0}^{d-1} m_\mu$.
A basis for the space of $p$-jets consists of all Taylor coefficients
$\phi_{,m}$ of order $|m|\leq p$, together with the expansion point $q$.
Unlike $q$, the point $x$ is a c-number which labels the field
components $\phi(x)$.

To illustrate how the Virasoro-like extensions of $\vect(d)$ arise,
it suffices to consider $-1$-jets, whose basis consists of the
expansion point only. $q^\mu$ and its conjugate momentum $p_\mu$
satisfy the canonical commutation relations
\be
[q^\mu, p_\nu] = i \delta^\mu_\nu.
\label{qp}
\ee
It follows immediately from the definition that the 
diffeomorphism generators on the torus admit the realization
\be
L_\mu(m) = \e^{im\cdot q} p_\mu.
\label{embed}
\ee
This equation defines an embedding of $\vect(d)$ into the universal
enveloping algebra of (\ref{qp}), and hence a represen\-tation of
$\vect(d)$ on the corresponding Fock space. 

Since the Heisenberg algebra (\ref{qp}) is finite-dimen\-sional, no
extension can arise from normal ordering. However, an extension does
arise with a slight modification of the construction. We consider 
one-dimen\-sional trajectories in the space of $-1$-jets, instead of 
just the $-1$-jets themselves. For technical simplicity, we consider
closed trajetories, i.e. circles, even though it may be physically
dubious to introduce closed time-like loops. The space of trajectories
has the basis $q^\mu(t)$, $p_\mu(t)$, $t \in S^1$, and satisfy the
canonical commutation relations
\be
[q^\mu(t), p_\nu(t')] = i \delta^\mu_\nu \delta(t-t').
\label{qpt}
\ee
The embedding of $\vect(d)$ into this algebra is completely analogous
to (\ref{embed}):
\be
L_\mu(m) = \int dt\ \e^{im\cdot q(t)} p_\mu(t).
\label{Lmu}
\ee
The quantization step consists of constructing lowest-energy 
represen\-tations of the Heisenberg algebra (\ref{qpt}). Since
$q^\mu(t)$ and $p_\mu(t)$ depend on a variable on the circle, they
can be divided into positive and negative frequency modes, which we
denote by $q_>^\mu(t)$, $q_<^\mu(t)$ and $p^>_\mu(t)$, $p^<_\mu(t)$, 
respectively; where the zero modes are assigned is not important.
The Fock vacuum is defined by
\be
q_<^\mu(t) \ket 0 = p^<_\mu(t) \ket 0 = 0.
\ee
The diffeomorphism generators (\ref{Lmu}) must be normal ordered to
act in a well-defined manner on the Fock vacuum. We define
\bes
L_\mu(m) &=& \int dt\ \no{ \e^{im\cdot q(t)} p_\mu(t) } 
\nlb{:Lmu:}
&=& \int dt\ \bigg( \e^{im\cdot q(t)} p^<_\mu(t)
+ p^>_\mu(t)\e^{im\cdot q(t)} \bigg).
\eens
By direct calculation, we find that the normal-ordered generators
(\ref{:Lmu:}) satisfy the Virasoro-like extension (\ref{mVir}) of
$\vect(d)$, where 
\be
S^\mu(m) = {1\/2\pi} \int dt\ \dot q^\mu(t) \e^{im\cdot q(t)},
\label{Smu}
\ee
and the parameters are $c_1 = 2d$, $c_2 = 0$.

Two observations are in order. First, the condition 
$m_\mu S^\mu(m) = 0$ is equivalent to demanding that integrals over
total derivatives vanish, i.e.
\be
\int dt\ {d\/dt}F(q(t)) \equiv 0,\qquad \hbox{for every function $F$.}
\ee
This condition is automatically satisfied because the integral runs
over a circle. Second, in one dimen\-sion there is only one circle; we can
therefore use the circle coordinate $q$ as the independent variable
rather than $t$, and the integral becomes
\be
S(m) = {1\/2\pi} \int dq\ \e^{im\cdot q} = \delta_m.
\ee
Hence (\ref{Smu}) reduces to the central Kronecker-delta extension of
the ordinary Virasoro algebra when $d = 1$.

More general represen\-tations, based on $p$-jets of tensor fields
instead of $-1$-jets, were constructed in \cite{Lar98}. The parameters
$c_1$ and $c_2$, as well as other ``abelian charges'', depend on the
truncation order $p$ as well as a $gl(d)$ represen\-tation, which
labels the type of tensor field that we start from. In contrast, the
operator $S^\mu(m)$ is always realized in the same way (\ref{Smu}).

This clearly shows why this kind of anomaly can never arise within the
framework of QFT. The extension is a functional of the expansion
points $q^\mu(t)$, which are never introduced in QFT. Hence it is
impossible to even write down the Virasoro-like anomalies within a QFT
framework. But it is also clear what the remedy is: work with the jets
instead of the fields, because a jet, or Taylor series, automatically
carries information about the expansion point.

\section{ QJT as a regularization }
\label{sec:reg}

This section closely follows the treatment in \cite{Lar09}. To extract
the physical content, it is useful to decompose spacetime into space
and time; boldface quantities ($\xx$, $\qq$, $\mm$) are used to denote
spatial components. For definiteness, we consider a free
scalar field with mass $\ww$. To the field $\phi(\xx,t)$ correspond
jet data $\phi_\cm(t)$, $\qq(t)$, related through its Taylor series:
\be
\phi(\xx,t) = \sum_\mm {1\/\mm!} \phi_\cm(t) (\xx - \qq(t))^\mm.
\label{Taylor}
\ee
Here $\mm = (m_1, ..., m_{d-1})$ is a spatial multi-index of length
$|\mm| = \sum_{j=1}^{d-1} m_j$, and all $m_j \geq 0$. 

Jets come equipped with a natural regularization: pass from 
$\infty$-jets to $p$-jets, i.e. truncate all Taylor series at order
$p$. This means that the sum in (\ref{Taylor}) only runs over
$\mm$ of length $|\mm| \leq p$. The correct equations of motion in 
$p$-jet space take the form $\EE_\cm(t) = 0$, where
\be
\EE_\cm = \begin{cases}
\phi_{,\mm+2\hat0} - \sum_{j=1}^{d-1} \phi_{,\mm+2\hatj} 
+ \ww^2\phi_\cm, & |\mm| \leq p-2 \\
\hbox{undefined} & |\mm| = p-1, p
\end{cases}
\label{Epjet}
\ee
Here $\mm+n\hatj = (m_1, ..., m_j + n, ..., m_{d-1})$ and
\be
\phi_\cmnoll \equiv \dot\phi_\cm - \sum_{j=1}^{d-1} \dot q^j \phi_\cmj.
\ee
The key observation is that $\EE_\cm$ is not defined for the
modes with $|\mm| = p-1, p$, i.e. the ``skin'' of the $p$-jet. The 
expression used for $|\mm|\leq p-2$ (the ``body'') would involve the 
components $\phi_{,\mm+2\mu}$ with $|\mm| > p$, which do not belong to 
$p$-jet space (they do belong to $(p+2)$-jet space). 
The full $p$-jet phase space, i.e. the space of $p$-jet histories
which solve the equations of motion (\ref{Epjet}), is hence 
spanned by
\bes
\phi_\cm(0), \pi_\cm(0) = \phi_\cmnoll(0), 
 &\qquad& |\mm| \leq p-2, \nle
\phi_\cm(t), \quad\quad \forall t \in \RR, 
 &\qquad& |\mm| = p-1, p.
\eens
The $p$-jet phase space is infinite-dimen\-sional because the equations
of motion are unable to determine some histories in terms of data 
living at $t=0$. 
This is the origin of the new gauge and diff anomalies. The ``body''
consists of only finitely many degrees of freedom, and can
hence not contribute to anomalies, but the ``skin'' is
infinite-dimen\-sional and normal ordering leads to anomalies in
theories with a gauge symmetry. In fact, we can even give the free
scalar field a gauge symmetry under reparametrizations of the 
observer's trajectory, by not identifying $x^0 = q^0(t)$ with the
parameter $t$. The algebra of reparametrizations then becomes a 
Virasoro algebra. The contribution from a single bosonic function of
$t$ to the central charge is $c = 2$. The number of different
multi-indices with $|\mm|\leq r$ in $d-1$ space dimen\-sions is 
${d+r-1 \choose d-1}$. The total central charge for the ``skin'' is
thus
\be
c_\Tot = 2{d+p-1 \choose d-1} - 2{d+p-3 \choose d-1}.
\label{ccBose}
\ee

\section{ Finite anomalies }
\label{sec:fin}

The passage to $p$-jets is a regularization, and in the end we want to
eliminate it by taking the limit $p\to\infty$. However, the expression
for $c_\Tot$ in (\ref{ccBose}) diverges in this limit, provided that
$d \geq 3$. Whereas finite gauge and diff anomalies may be consistent,
an {\em infinite} anomaly is certainly a sign of inconsistency.
Fortunately, in \cite{Lar04} I discovered a way to avoid this
problem.

Consider a theory with fermions with central charge $c_F$, 
with bosons with central charge $-c_B$, and with gauge fields
with central charge $-c_G$. The total central charge is given by
\bes
c_\Tot &=& (c_F - c_B) {d+p-1 \choose d-1}
- c_F {d+p-2 \choose d-1} 
\nlb{cTot}
&&+\ (c_B + c_G) {d+p-3 \choose d-1}
- c_G {d+p-4 \choose d-1}.
\eens 
If we choose the field content such that
\be
c_F = 3c, \qquad c_B = 2c, \qquad c_G = c,
\ee
for some $c > 0$, the total central charge (\ref{cTot}) reduces to
\be
c_\Tot = c {d+p-4 \choose d-4}.
\ee
In particular, $c_\Tot = c$ is independent of $p$ if $d = 4$, and
the infinite parts of the reparametrization anomaly have been
cancelled.

Hence, if we choose a theory with a natural field content (fermions
with first-order equations of motion, bosons with second-order
equations of motion, and with gauge symmetries which are not
reducible), we can cancel the infinite part (but not the finite part)
of the reparametrization anomaly iff spacetime has four dimen\-sions.
The same is true for the infinite parts of gauge and diff
anomalies \cite{Lar04}. This strongly suggests that spacetime has four
dimen\-sions.

Unfortunately, making a more detailed identification of the field
content with expermentally existing matter leads to problems which
remain open.

\section{ Conclusion }

We have reviewed key features of Quantum Jet Theory: observer
dependence and the multi-dimen\-sional Virasoro algebra. The appearence
of new diff anomalies shows that QJT is substantially different from
QFT, which is positive given that QFT is incompatible with gravity.
Virasoro-like diff anomalies may well be consistent, and are in fact
a necessary ingredient in any local, non-holographic, quantum theory
of gravity.

\end{document}